\documentclass[a4paper]{jpconf}
\usepackage{graphicx,cite}
\begin{document}
\title{Hot subdwarf stars and their connection to thermonuclear supernovae}

\author{S. Geier$^{1,2}$, T. Kupfer$^3$, E. Ziegerer$^2$, U. Heber$^2$, P.~N\'{e}meth$^{2}$, A. Irrgang$^{2}$ and MUCHFUSS team}

\address{$^1$ Department of Physics, University of Warwick, Coventry CV4 7AL, UK\\
$^2$ Dr. Karl Remeis-Observatory \& ECAP, Astronomical Institute, Friedrich-Alexander University Erlangen-Nuremberg, Sternwartstr. 7, D~96049 Bamberg, Germany\\
$^3$Division of Physics, Mathematics, and Astronomy, California Institute of Technology, Pasadena, CA 91125, USA}

\ead{$^1$s.geier@warwick.ac.uk}

\begin{abstract}
Hot subdwarf stars (sdO/Bs) are evolved core helium-burning stars with very thin hydrogen envelopes, which can be formed by common envelope ejection. Close sdB binaries with massive white dwarf (WD) companions are potential progenitors of thermonuclear supernovae type Ia (SN~Ia). We discovered such a progenitor candidate as well as a candidate for a surviving companion star, which escapes from the Galaxy. More candidates for both types of objects have been found by crossmatching known sdB stars with proper motion and light curve catalogues. The Gaia mission will provide accurate astrometry and light curves of all the stars in our hot subdwarf sample and will allow us to compile a much larger all-sky catalogue of those stars. In this way we expect to find hundreds of progenitor binaries and ejected companions.
\end{abstract}

\section{Introduction}
Hot subdwarf stars (sdO/Bs) are evolved core helium-burning stars with very thin hydrogen envelopes, which can be formed by common envelope ejection. Close sdB binaries with massive C/O-WD companions are candidates for supernova type Ia (SN~Ia) progenitors, because mass-transfer can lead to the thermonuclear explosion of the WD. The project Massive Unseen Companions to Hot Faint Underluminous Stars from SDSS (MUCHFUSS) aims at finding the sdB binaries with the most massive compact companions like massive white dwarfs, neutron stars or black holes.

We selected and classified about $\sim1400$ hot subdwarf stars from the Sloan Digital Sky Survey (SDSS DR7). Stars with high velocity variations have been reobserved and analysed. In total $177$ radial velocity variable subdwarfs have been dis\-covered and $1914$ individual radial velocities measured. We constrain the fraction of close massive companions of H-rich hot subdwarfs to be smaller than $\sim1.3\%$ \cite{geier15b}. Orbital parameters as well as minimum companion masses have been derived from the radial velocity curves of 30 sdB binaries \cite{kupfer15}. 

\section{Discovery of an SN\,Ia progenitor and an ejected companion}

We detected high RV-variability of the bright sdB CD$-$30$^\circ$11223. Photometric follow-up revealed both shallow transits and eclipses, allowing us to determine its component masses and fundamental parameters. The binary system, which is composed of a C/O-WD ($\sim0.76\,M_{\rm \odot}$) and an sdB ($\sim0.51\,M_{\rm \odot}$) has a very short orbital period of $\sim0.049\,{\rm d}$. In the future mass will be transfered from the helium star to the white dwarf. After a critical amount of helium is deposited on the surface of the white dwarf, the helium is ignited. Modelling this process shows that the detonation in the accreted helium layer should be sufficiently strong to trigger the explosion of the core. Thermonuclear supernovae have been proposed to originate from this so-called double-detonation of a WD \cite{fink10,geier13}. The surviving companion star will then be ejected with its orbital velocity. The properties of such a remnant match the hypervelocity star US\,708, a helium-rich sdO star moving with $\sim1200\,{\rm km\,s^{-1}}$, exceeding the escape velocity of our Galaxy by far and making it the fastest unbound star known in our Galaxy \cite{geier15a}.

\section{Finding more progenitor and ejected companion candidates}

Since the properties of the ejected companions, especially the ejection velocity, allow us to constrain the properties of the binary progenitors such as the orbital period and the companion mass right at  the moment of the explosion, we will gain an unprecedented insight into the formation of SN\,Ia and learn about other acceleration mechanisms for hypervelocity stars, if more such objects can be found and studied. The distribution of orbital periods and WD companion masses of progenitor binaries will help us to constrain SN\,Ia progenitor models. While binaries with periods longer than about 2\,hr will merge as double degenerates, closer binaries might be progenitors for the helium double-detonation channel.

To search for ejected companions we compiled a catalogue of all known sdO/B stars from the literature and our own database ($\sim4500$ stars) and crossmatch it with proper motion catalogues. Candidates with high velocities are followed-up with spectroscopy (Keck/ESI, VLT/XSHOOTER, SOAR/Goodman, CAHA/TWIN, WHT/ISIS) to measure spectroscopic distances and derive kinematics. Several good candidates for unbound hypervelocity sdO/Bs have already been found.

We found that hot subdwarf binaries with massive WDs in close orbits are quite rare. To find more of those objects, we crossmatch the hot subdwarf catalogue with light curve catalogues (e.g. CRTS, PTF, SWASP, GALEX gPhoton, Kepler K2) and search for the characteristic sinusoidal variations caused by the ellipsoidal deformation of the sdB. Several candidates have been found and will be followed-up spectroscopically and photometrically.

Eventually, the Gaia mission will provide accurate astrometry and light curves of all the stars in our hot subdwarf sample and will allow us to compile a much larger all-sky catalogue of those stars. In this way we expect to find hundreds of progenitor binaries and ejected companions.\\\\

\bibliography{geier_proc}

\end{document}